\documentclass[reprint,prl,a4paper,nofootinbib,showpacs, showkeys,
twoside]{revtex4-1} 
\usepackage[T2A]{fontenc}
\usepackage[english]{babel} 
\usepackage{amsthm,graphicx,color}
\usepackage[psamsfonts]{amssymb}
\usepackage{mathrsfs}

\usepackage{hyperref}
\hypersetup{pdfauthor={Brezhnev},
						colorlinks=true, citecolor=blue,
						pagebackref=true, 
						pdfwindowui=false, pdfmenubar=false, 
						bookmarksopen=false, bookmarks=false, linktocpage=false,
						hyperfootnotes=false}

\usepackage{YURA,Brackets,mScripts,widebar,Ulem}

\renewcommand{\,}[1][1]{
\ifmmode\mkern#1mu\else\kern#1\dimexpr0.05556em\fi\relax}
\renewcommand{\!}[1][1]{\,[-#1]}

\makeatletter\def\@makefnmark{\textup{\red*}}\makeatother

\theoremstyle{remark}%
\newtheorem{comm}{
\indent \iflanguage{russian}{Замечание}{Remark}~~\ignorespaces}
\newenvironment{comment}%
{\vspace{-1ex}\par\smaller[0.5]\begin{comm}}{\end{comm}\vspace{-1ex}}

\def\SL{\textup{\texttt{Stat\-Length}}}

\def\+{\mathbin{\hbox to0ex{\kern0.05556em\Under[-0.1]%
{\phantom{\mkern-2mu+}}}\raise0.1ex\hbox{$+$}}}

\def\h+{\mathbin{\hat+}}

\newcommand{\TRANS}[2][2]{%
\mathrel{\mathop{{\mathrel{%
\REPEAT{\mathchar"439\mkern-1mu}{#1}\mkern-7mu\rightsquigarrow}}}%
\limits^{\raise-1ex\hbox{\smash{$#2$}}}}}

\newcommand{\GOTO}[2][2]{\mathrel{\mathop{\goto[#1]}%
\limits^{\raise-1ex\hbox{\smash{$#2$}}}}}

\def\state#1{\underline{#1}{}}
\def\ket#1{|\bo#1\rangle}

\def\obj#1{\raise0.4ex\hbox{$\sss\mkern-1mu\left\lgroup\right.
\mkern-4mu$}#1\raise0.4ex\hbox{$\sss\mkern-2.5mu\left.
\right\rgroup\mkern-4mu$}}

\def\ketPsi{\ket{\Psi}}
\def\ketPhi{\ket{\Phi}}

\begin{document}

\title{The Born rule}
\author{\itsf{Yurii V.~Brezhnev}}
\affiliation{Department of Quantum Field Theory, Laboratory of
mathematical physics, Tomsk State University, Russia}



\begin{abstract}
We deduce the Born rule. No use is required of quantum postulates.
One exploits only rudimentary quantum mathematics---a linear, not
Hilbert', vector space---and empirical notion of the statistical
length of a state. Its statistical nature comes from the
experimental detector"=clicks being formalized into the abstract
quantum micro"=events. We also comment on that it is not only that
the use has not been made of some quantum axioms when deriving the
rule but, in a sense, their invoking would be inconsistent.
%
\end{abstract}
\maketitle

The connection between quantum theory and physical experiment begins
with the famous `square of the magnitude' formula $|\psi|^2$. This
intuitive guess by Max~Born \cite{born}---``an intuition without a
precise justification'' (A.~Cabello)---determines the statistical
interpretation of quantum wave"=function and, presently, no violation
of the `rule of squares' has ever been discovered. Its purposeful
experimental testing, however, came into implementation relatively
recently. The pioneering works of U.~Sinha et al \cite{sinha,
sinha+} have demonstrated, in a 3-slit interference laser
experiment, the null"=effect within an accuracy $10^{-2}\pm10^{-3}$.
The rule is considered as one of the cornerstone of the theory,
although many researchers have long pointed out \cite{ballentine,
everett, deutsch, hartle, saunders1, zurek1, zurek2, zurek3,
saunders}, and it seems to be a majority opinion, that this Born
formula stands apart from other tenets of quantum mechanics (\qm),
because it can be derived from the other ones and, thereby, it is
not a fundamental `mantra'. Attempts at deriving the rule are of
great variety, reveal interesting parallels \cite{saunders1}, and
have been the subject of an extensive literature.

Gleason \cite{gleason} 
proved an advanced version of Born's result as a statement about
abstract measures on Hilbert's spaces (see also \cite{busch} for a
more comprehensive variant of this result), and Everett, in the
framework of his famous treatment of \qm\ \cite{saunders},
considered specifically the rule \cite[pp.~71--72]{everett}. In
1999, D.~Deutsch \cite{deutsch} revived Everett--DeWitt's approach
and initiated a new one, which relates the \qm"=theory with the
representation theorems of classical decision theory through the
characteristic terminology: strategies of a rational agent, bets,
weight/""utility functions (attributed to experimental outcomes), game
theory, \etc. Deutsch's ideas were refined by Wallace
\cite{wallace}, \cite[p.~227--263]{saunders} and Saunders
\cite{saunders1} in the 2000's; see also \cite{assis}, Ch.~3 in the
book \cite{saunders} and bibliography therein. W.~Zurek, by the
``fine/""coarse"=graining'' technique \cite{zurek1, zurek2, zurek3},
developed a different---envariance/""decoherence---strategy for
deriving the rule. Graham \cite{graham} and Hartle \cite{hartle}, in
the 1960--70's, have proposed the frequency"=operator method. There
are other ways of looking at the problem \cite[Ch.~5--6]{saunders},
\cite{aaronson,zela, masanes}. These references are by no means
complete; say, the \texttt{arXiv}"=search yields hundreds items with
mentioning the `Born rule' in abstracts.

All approaches---for extended bibliography see
\cite{mandolesi}---have been subject to mutual criticism
\cite[Ch.~4]{saunders}, \cite[p.~25]{zurek2}, \cite{rae, caves,
ballentine, squires, jarlskog, mandolesi}. In particular, most if
not all of the derivations appeal to unitary $t$"=evolution and
tensor products, whereas neither of these concepts has been present
in Gleason's theorem. One of the typical objections voiced against
the alternative ideas is circular reasoning \cite{saunders,
mandolesi, caves, zurek3, graham, dewitt}. This is a criticism made
not just by proponents of one approach towards another, but one that
is sometimes admitted by the authors of the ideas themselves
\cite[p.~415]{saunders}. Most of the known approaches, including the
Everettian one, have undergone revisions and refinements
\cite[Ch.~5]{saunders}, \cite{zurek3, dewitt}. These points reflect
the long"=standing problem with quantum foundations---linguistic
self"=referentiality in their substantiating. Thus the situation
seems to be one whereby the numerous attempts to rationalize the
`square' preserve the status quo; none of the approaches have been
widely accepted to date. The Born rule is continuing to exist as an
arduous task, especially considering that the formula should be
derived rather than being proved.

In this work we exhibit a straightforward deducing the mod"=squared
dependence. In doing so, it is suffice to rely not on the canonical
axiomatics (of a Hilbert space) but on a formulation of
\qm"=foundations as a theory of micro"=events/""clicks \cite{br}. One
uses only the most primitive property of the quantum"=state set: to
be a linear vector space (\lvs). The primary idea of
derivation---separation of the number entities---was in effect
announced in sects.~9.1--2 of the work \cite{br}. These sections,
including some illustrative counterexamples therein, can be
considered as an extended introduction to the present work and we
reproduce the ideology here very briefly.

\vbox{\section{Doctrine of numbers\\in quantum theory, revisited}}

Theory begins with a number, and intuitive perception of this object
is always accompanied by the notion of a physical unit
\cite[sects.~7.1--3 and Remark~16]{br}. This is the interpretative
reading the number in terms of the `quantity of something real':
metres, St\"ucke, sheep, \etc. The releasing the number from such
units---mathematization---turns it into an abstract operator
$\oper[-2]n$ and, then, into an abstract element $\msf n$ of the
abstract set $\bbR$ with arithmetic operations $\{+,\times\}$.
Thereupon there arises a $\bbC$"=structure of the complex numbers
$\frak a\DEF\msf n+\ri\,\msf m$ equipped with the binary operations
$\{\oplus,\odot\}$ and unary involutions
\begin{equation}\label{inv}
(\msf n+\ri\,\msf m)\overset{\raisebox{-0.21ex}{$\scs\smash*$}}
\mapsto(\msf n-\ri\,\msf m)\,,\qquad (\msf n+\ri\,\msf m)\mathbin{%
\overset{\raisebox{-1ex}{\smash{\textasciitilde}}}%
{\smash\mapsto}}(\msf m+\ri\,\msf n)\,.
\end{equation}
Recall \cite[\textbf{\textsf{II}}-nd principium of \qm]{br}, with
specification of the number conceptions missing, the exegesis of
`everything the quantum' acquires the character of a circular
argument. The last step is a creation of the other kind
number"=entities: \emph{non}"=abstract, reified quantities per~se. It
constitutes a mathematical realization of what we have been calling
observable quantities: statistics, spectra, means, \etc. In a word,
the abstract numbers and the observational ones are significantly
distinct by their nature. What is more, the quantum foundations
themselves may not be grounded on the physical/""observable concepts
and their numeric forms.

Now, the quantum mathematics, in its rudimentary form, is \emph{but
an abstract\/\footnote{The abstracta themselves, the process of
abstracting, its naturalness and inevitability are the subject
matter of a comprehensive discussion in sects.~9.2--3 of \cite{br}.}
algebra} of a linear space $\bbH$ over the $\tilde\bbC^*$"=number
objects and $\ket\alpha$"=expansions
\begin{equation}\label{=}
\frak a_1\bo\cdot\ket{\alpha_1}\h+
\frak a_2\bo\cdot \ket{\alpha_2}\h+\cdots\quad\in\bbH
\end{equation}
with respect to the eigen"=vectors $\ket{\alpha_j}$ of an instrument
$\scr A$. Because the notion of `observable' is initially absent not
only in nature but in theory as well,
\begin{itemize}
\item[\hypertarget{I}{($\red\bullet$)}] the numerical\label{I} values of that
 which is associated with the term `observable quantity' may arise
 only as a supplement to the $\bbH$"=algebra: the extra rules for
 manipulating the symbols
 $\{\frak a,\ket\alpha,\bo\cdot\,,\h+,\oplus,\odot,*,
 \raisebox{-0.9ex}{\textasciitilde}\}$ in the
 construct~\eqref{inv}--\eqref{=}.
\end{itemize}

These rules must constitute the mathematical maps---math add-ons
over $\bbH$---into the ordered continuum equipped with arithmetic
$\{+,\times\}$; the $\bbR$"=numbers for short. The `ordered' here
because of the language's notion `greater/""lesser'. Such a scheme
delivers the only means of formalizing anything that accompanies the
low"=level quantum mathematics in the form of notions that we portray
in terms of natural language. These are usually referred to as
physical quantities.

For example, statistics of $\state\alpha$"=clicks
\cite[sect.~2.5--6]{br}, \ie, the relative frequencies
$(\nu_1,\nu_2,\ldots)$ may come only from $\frak a$"=coefficients
in~\eqref{=}:
\begin{equation}\label{nu}
(\frak a_1,\frak a_2,\ldots)\mapsto\nu_j\quad\hhence\quad
\nu_j=\smash{\overset{\smash{(?)}}{f_j}}(\frak a_1,\frak a_2,\ldots)\,.
\end{equation}
However, the \emph{$\nu$"=numbers are not the primordial empirical
entities}. In experiments---colliders, ion traps,
inter\-fe\-ro\-me\-ters, or any other quantum installation---we are
dealing not with quantities that are subject to `rather specific'
constraints $0\leqslant\nu_j\leqslant1$---a theoretical act that does not follow
from \qm"=empiricism---but with gathering the registered micro"=events
$\state\alpha_j$. It has been just these (additive) accumulations,
being formalized into $\frak a_j$"=coefficients, which are to be
turned into the $\bbR$"=numbers mentioned above, because it is in
this way that the number tokens arise in theory at all
\cite[sect.~7.2]{br}. Therefore, what is taken as a primary
mathematical map must be not \eqref{nu} but what we shall call
\emph{statistical length of an $\ket\alpha$"=representation}:
\begin{equation*}
\text{\SL\ of \eqref{=}}\,.
\end{equation*}

Inasmuch as mathematics of $\state\alpha$"=clicks implies the
infiniteness of quantum-click ensembles, the \SL\ should be created
as a mathematical equivalent to the empirical wording `the quantity
of micro"=events' having regard to---also
empirical---$\Sigma$"=postulate about theoretical infinity of the
event number \cite[sect.~2.5]{br}. Considering that ``das Unendliche
findet sich nirgends realisiert'' (D.~Hilbert (1926)), we rely on the
following underlying semantics:
\begin{equation}\label{infty}
\text{(infinite) number of $\state\alpha$-clicks} = \SL\times
\infty\,.
\end{equation}
Notice that the integer-valued domain $\bb Z$, as such, does not
appear in quantum theory. The discrete infinity $\aleph_0$, upon
applying the $\Sigma$"=postulate, disappears and yields to continuum
$2^{\aleph_0}$. That is, an infinite accumulation of the
$\lceil\text{event"=number}\rightarrowtail\bb Z\rceil$ leads to the
following sequence of infinities
\begin{equation*}
(\bb Z\times\infty) \rightarrowtail\aleph_0 \rightarrowtail
2^{\aleph_0}\rightarrowtail\bbR \rightarrowtail(\bbR\times\bbR)\FED
\bbC\overset{\eqref{nu}\,}\mapsto\bbR^{\sss+}
\end{equation*}
\cite[sect.~4]{br}. It is the quantum ensembles that give birth to
the state"=vector $\frak a\bo\cdot\ketPsi$ itself. The function \SL\
is thus understood further to be the $\bbR^{\sss+}$"=numeric one. Let
us take a closer look at the situation, in order to ascertain
properties of this function.

\section{Axioms of statistical length}

First and foremost, the \SL\ is associated only with \eqref{=}
because quantum ``empiricism \ldots\ yields \emph{not states} and
superpositions thereof \emph{but $\ket\alpha$"=representations}''
\cite[sect.~8.3]{br}. It is such representations that are primary in
\qm\ rather than the formal (superpositions of) states. For example,
the writing $\SL\,(\ketPsi\h+\ketPhi)$ lacks meaning---or rather, in
no way determinable---unless the $\ketPsi$ and $\ketPhi$ are
indicators of certain eigen"=elements. At the same time, the writing
$\SL\,(\ketPsi)$ is admissible since any element
$\ketPsi=1\bo\cdot\ketPsi\in\bbH$ may serve as the eigen one for a
certain instrument~$\scr B$. What are the empirical definienda
(linguistic semantics) for the conception \SL?

Each of $\state\alpha$"=clicks, in accord with their
($\approx$)"=distinguishability, corresponds to a certain ket
$\ket{\alpha_j}\leftrightarrow\state\alpha_j$. Consequently, the
need for frequencies \eqref{nu} means that the partial lengths
$\SL\,(\frak a_j\bo\cdot\ket{\alpha_j})$ should come into play.
Certainly, these lengths must correlate with the total \SL\ of
\eqref{=}. Besides, the numeric values of all the \SL's appear to be
compatible with each other, for any statistical
$\scr A$"=representative
\begin{equation*}
\frak a_1\bo\cdot\ket{\alpha_1}\h+
\frak a_2\bo\cdot\ket{\alpha_2}\h+\cdots=\cdots
\end{equation*}
is re"=developable with respect to other instrument $\scr B$:
\begin{equation}\label{AB}
\cdots=\frak b_1\bo\cdot\ket{\beta_1}\h+
\frak b_2\bo\cdot\ket{\beta_2}\h+\cdots\,.
\end{equation}
What is more, even the very formal $\ketPsi$"=object cannot be
constructed without matching the two instruments $\scr A$, $\scr B$
\cite[sect.~5.4]{br}. Let us agree to call the relation
\begin{equation*}
\SL\,\big(\frak a_1{\bo\cdot}\ket{\alpha_1} \h+\cdots)=
\SL\,\big(\frak b_1{\bo\cdot}\ket{\beta_1} \h+\cdots)
\end{equation*}
the instrument- or device"=independence.

If the two events $\state\alpha_1$ and $\state\alpha_2$ are
distinguishable by the $\scr A$"=instrument
($\state\alpha_1\not\approx\state\alpha_2$) then the statistical
length of a $(\h+)$"=sum of two statistical
$\ket\alpha$"=representatives
\begin{equation*}
\SL\,\big(\frak a_1{\bo\cdot}\ket{\alpha_1} \h+ \frak a_2{\bo\cdot}
\ket{\alpha_2}\big)=\cdots\,,
\end{equation*}
by the very nature of `the number of clicks' and of `the mutual
exclusivity of $\state\alpha$'s', is split into the numeric sum of
the partial lengths:
\begin{equation}\label{add}
\![3]\cdots=\SL\,\big(\frak a_1\bo\cdot\ket{\alpha_1}\big)\,+\,\,
\SL\,\big(\frak a_2\bo\cdot \ket{\alpha_2}\big)\,.
\end{equation}
This property determines a translation (homomorphism) of the
`abstract' $(\h+)$"=operation on $\bbH$"=vectors into the `concrete
arithmetical plus $+$' between the $\bbR$"=numbers. Of course, this
is a peculiarity of the $\ket\alpha$"=bases, not of the arbitrary
ones.

Meantime, there is yet another operation with the
$\bbH$"=vectors---the unary multiplication $\ket\alpha\mapsto
\frak c\bo\cdot\ket\alpha$---and it should also be carried over to
the arithmetic of the \SL"=numbers:
\begin{equation*}
\smash{\SL\,(\ket\alpha)
\mathbin{\overset{{\skew{2}\widehat{\ds\frak c}\,\,}}{\mapsto}}
\SL\,(\frak c\bo\cdot\ket\alpha)={}?}
\end{equation*}
Clearly, $\SL\,(\frak c\bo\cdot\ket\alpha)$ is a certain function of
the \SL\ of $\ket\alpha$. Therefore, simplifying notation
$\SL\,(\cdot{\cdot}\cdot)\rightarrowtail\cal N[\cdot{\cdot}\cdot]$,
we have to find a $\textsf{C}$"=function:
\begin{equation}\label{times}
\cal N[\frak c\bo\cdot\ket\alpha]=
\overset{\smash{(?)}}%
{\textsf{C}}\!\!\big(\cal N[\ket\alpha]\big)\,.
\end{equation}

On the other part, $\ket\alpha$"=objects are elements of \lvs. This
means that the $\cal N$"=function must respect its axioms. In
particular, the distributivity
\begin{equation}\label{axiom}
\makebox[0ex][r]{$\frak c\bo\cdot(\ket\alpha\h+
\ket\beta)$}=
\makebox[0ex][l]{$\frak c\bo\cdot\ket\alpha\h+
\frak c\bo\cdot\ket\beta$}
\end{equation}
entails
\begin{equation*}
\makebox[0ex][r]{$\cal N\big[\frak c\bo\cdot(\ket\alpha\h+
\ket\beta)\big]$}=
\makebox[0ex][l]{$\cal N\big[\frak c\bo\cdot\ket\alpha\h+
\frak c\bo\cdot\ket\beta\big]\,.$}\notag
\end{equation*}
When $\ket\alpha$ and $\ket\beta$ correspond to distinguishable
clicks $\state\alpha\not\approx\state\beta$, the additivity
\eqref{add} entails a translation $(\h+)\mapsto(+)$ on the right:
\begin{align*}
\cal N\big[\frak c\bo\cdot(\ket\alpha\h+\ket\beta)\big]&=
\cal N[\frak c\bo\cdot\ket\alpha]+
\cal N[\frak c\bo\cdot\ket\beta]\,.
\intertext{All the $\cal N$"=functions here are the ones of
$[\frak c\bo\cdot(\cdot{\cdot}\cdot)]$. Hence,}
\textsf{C}\big(\cal N[\ket\alpha\h+\ket\beta]\big)&=
\textsf{C}\big(\cal N[\ket\alpha]\big)+
\textsf{C}\big(\cal N[\ket\beta]\big)
\intertext{and, applying additivity \eqref{add}, now on the left, we
obtain}
\textsf{C}\big(\cal N[\ket\alpha]+
\cal N[\ket\beta]\big)&=
\textsf{C}\big(\cal N[\ket\alpha]\big)+
\textsf{C}\big(\cal N[\ket\beta]\big)\,.
\end{align*}
The $(\h+)$"=abstractum disappears and we arrive at the standard
functional equation for the linear (real"=valued, continuous) numeric
function \cite[pp.~128--129]{kuzma}:
\begin{equation}\label{Cx}
\textsf{C}(x+y)=\textsf{C}(x)+\textsf{C}(y)\quad\hence\quad
\textsf{C}(x)=\msf{const}\times x\,.
\end{equation}
Thus, the abstract ($\bo\cdot$)"=sign in \eqref{times} has been
converted into the numerical $\times$. Summing up, we introduce the
function $\cal N$ by a definition, which will suffice to derive the
rule.

\begin{widetext}
\textbf{Definition} (\emph{axioms of \SL}). %
The $\bbR$"=valued function $\cal N$ formalizes
\textup{(homomorphically)} the statistical"=length conception by the
rules of carrying the abstracta $\{\h+,\bo\cdot\}$ over to the
arithmetic $\{+,\times\}$\,:
\begin{Align}
(\h+)&\mapsto(+)&&:&\qquad\cal N\big[\frak a_1\bo\cdot\ket{\alpha_1}
\h+ \frak a_2\bo\cdot\ket{\alpha_2}\h+\cdots\big]
&{}={}\,\cal N[\frak a_1\bo\cdot\ket{\alpha_1}]+
\cal N[\frak a_2\bo\cdot\ket{\alpha_2}]+\cdots\quad&&
\forall\frak a_j\,,\label{N+}\\[1ex]
(\bo\cdot)&\mapsto(\times)&&:&\qquad
\cal N\big[\frak c\bo\cdot(\frak a\bo\cdot\ket\alpha)\big]&{}={}
\msf{const}\,(\frak c)\times\cal N[\frak a\bo\cdot\ket\alpha]&&
\forall\frak a,\frak c\,.
\label{scale}
\end{Align}
The total \SL\ is device"=independent (meaningfulness of the
\SL"=number)\/$:$
\begin{gather}
\begin{array}{c}
\makebox[0ex][r]{$\ds\frak a_1\bo\cdot\ket{\alpha_1}
\h+ \frak a_2\bo\cdot\ket{\alpha_2}\h+\cdots$} =
\makebox[0ex][l]{$\ds\frak b_1\bo\cdot\ket{\beta_1}\h+
\frak b_2\bo\cdot\ket{\beta_2}\h+\cdots$}\\[1ex]
\Downarrow\label{noncom}
\end{array}\\
\makebox[0ex][r]{$\ds\cal N\big[\frak a_1\bo\cdot\ket{\alpha_1}
\h+ \frak a_2\bo\cdot\ket{\alpha_2}\h+\cdots\big]$} =
\makebox[0ex][l]{$\ds\,\cal N\big[\frak b_1\bo\cdot\ket{\beta_1}\h+
\frak b_2\bo\cdot\ket{\beta_2}\h+\cdots\big]\,,$}\label{star}
\end{gather}
and the function $\cal N$ is invariant under involutions
\eqref{inv}:
\begin{equation}\label{Ninv}
\cal N[\frak a^*\bo\cdot\ket\alpha]=
\cal N[\frak a\bo\cdot\ket\alpha]=
\cal N[\tilde{\frak a}\bo\cdot\ket\alpha]\,.
\end{equation}
\end{widetext}

Property \eqref{scale} is actually not an axiom because the
sequencing between formulas \eqref{times} and \eqref{Cx} is a
derivation of \eqref{scale}; nor is \eqref{Ninv} an axiom \cite{br}.
Beyond that, the $(\times)$"=scalability \eqref{scale} may be
postulated even purely semantically. Indeed, an operator
characterization of the number \cite[sect.~7.2]{br}---no matter, of
the real/""complex---entails the replication of quantum ensembles. The
replication means that the quantity $\SL\,(\ketPsi)$, upon action of
the `$\oper{\frak c}$"=operator' on $\ketPsi$, is merely multiplied
by a certain $\bbR$"=$\msf{const}\,(\frak c)$. Speaking more loosely,
we are dealing with a kind of homomorphism
\begin{equation*}
({\cdot}\widehat\cdot{\cdot})\text{-replication}=
\Big\{\begin{array}{c}
\text{to be multiplied by $\cdots$}\\
\hence\quad\text{scalability \eqref{scale}}
\end{array}\Big\}\,.
\end{equation*}
But it is just this mechanism---a group with $\frak c$"=scalars as
operator automorphisms---that is realized in the axiomatic structure
which has been calling `\emph{the \lvs}'; in particular, axiom
\eqref{axiom}. See a selected thesis following Remark~16 in the work
\cite{br}. With regard to axiom \eqref{N+}, symbols $\h+$ and $+$
are inherited from the `ensemble"=accumulation theory' by means of
the union operation $\cup$ \cite[sect.~5]{br}.

A shorter way to put the said above is that the language usage of
the notion \eqref{infty}, in all the linguistic diversity of the
\SL's descriptions, will boil down to the formal precepts
\eqref{N+}--\eqref{Ninv}.

\begin{comment}
Let us return once again to the thesis~(\hyperlink{I}{$\bullet$}). The
$\ket\alpha$"=additivity, scalability, and device"=independence are
not merely semantic characteristics. That is, the meaning associated
with the terms `observable value/""lengths/""volumes \etc' \emph{is not}
something that is conceived of or rephrased by various words, but
precisely---and this we stress with emphasis---what's being
(abstractly) added ($+$), multiplied ($\times$), and calculated with
$\bbR$"=characters, irrespective of how it is being observed.
However, any math"=realization of this entity does not and cannot
exist a priori as a formula\footnote{``\emph{There is no arithmetic
in interferometers/""colliders}---there are only clicks there \ldots''
\cite[sect.~2.3]{br}.}. The latter is to be created from scratch
while we have no (more primary) mathematics at our disposal apart
from the $\bbH$"=space algebra: the $\tilde{\bbC}^*$"=numbers and
\lvs. Accordingly, \emph{there is no room here for
interpretations}---the correspondence
$\lceil\text{math}\rightarrowtail\text{phys}\rceil$---of
mathematical symbols in terms of (yet unclear) `observable
categories' or other words. The former are created on the basis of
the rules listed above and of the quantum-clicks' theory \cite{br}.
Or, if it comes to that, \emph{the} interpretation \emph{is in
itself the} rules \eqref{N+}--\eqref{Ninv}. No other sources of the
\SL"=formula exist.
\end{comment}

\section{The rule}

Now, we have to find the numeric $N(\frak a)$"=representation
\begin{gather}\label{Na}
\cal N[\frak a\bo\cdot\ket\alpha]=
\smash{\overset{{\,\,(?)}}{N}(\frak a)}\,.
\end{gather}
Stated differently, additivity \eqref{N+} creates the function $N$
(of a \emph{single numeric} argument) whose properties are specified
by \textbf{Definition}. The further strategy is to process axioms
\eqref{scale}--\eqref{Ninv}. The first step is \eqref{Ninv} and
scalability \eqref{scale}; the result will be
$N(\frak a)\sim|\frak a|^{2\cdot\textsf{p}}$. The second step concerns
the `arrow` \eqref{noncom}, which will result $N(\frak a) \sim
|\frak a|^{2\cdot1}$.

Condition \eqref{Ninv} tells us that $N(\frak a)$ must be invariant
upon actions of the non-$\bbC$"=algebraical involutions \eqref{inv}:
\begin{equation*}
N(\frak a^*)=N(\frak a)=N(\tilde{\frak a})\,,
\end{equation*}
but due to the algebraic connection
$\tilde{\frak a}=\ri\odot\frak a^*$, we may forget either of them.
Hence $N(\frak a)$ is a symmetrical function
$N_{\!*}(\frak a,\frak a^*)$ of the two ($\bbC$"=algebraically
independent) variables $(\frak a,\frak a^*)$ and can be represented
as an expression in the symmetrical polynomials
$\{1,\frak a\oplus\frak a^*,\frak a\odot\frak a^*\}$:
\begin{align}
N(\frak a)&=N_{\!*}(\frak a,\widebar{\frak a})\notag\\&
=1\,\gamma_0+\gamma_1\,(\frak a+\widebar{\frak a})+\gamma_2\,
(\frak a\,\widebar{\frak a})+ \gamma_3\,(\frak a+\widebar{\frak a})\,
(\frak a\,\widebar{\frak a})\quad\notag\\&\quad
+\cdots+ \gamma_{\ell p}\,(\frak a+\widebar{\frak a})^\ell\,
(\frak a\,\widebar{\frak a})^p+\cdots\label{N*}
\end{align}
($\gamma\text{'s}\in\bbR$, $\gamma={}?$). Here, as always in the
sequel, we have adopted a bar notation for the complex conjugation
$\frak a^*\FED\widebar{\frak a}$ and the stand\-ard convention for
the addition/""multiplication sym\-bols $\{\oplus,\odot\}$ and
$\{+,\times\}$ between both the $\bbC$- and $\bbR$"=numbers.

Let us consider the $N_{\!*}$"=representation of the scalability
property \eqref{scale}:
\begin{equation}\label{Ncc}
N_{\!*}(\frak c\,\frak a,\widebar{\frak c}\,\widebar{\frak a})=
\msf{const}\,(\frak c) \times N_{\!*}(\frak a,\widebar{\frak a})\,.
\end{equation}
This identity, upon substitution \eqref{N*}, reads as follows
\begin{multline*}
\smash{\sum_{\ell, p}}\gamma_{\ell
p}\,(\frak c\,\frak a+\widebar{\frak c}\widebar{\frak a})^\ell\,
(\frak c\,\frak a\,\widebar{\frak c}\widebar{\frak a})^p \\
=\msf{const}\,(\frak c)\,\times\smash[b]{\sum_{\ell, p}}\gamma_{\ell
p}\,(\frak a+\widebar{\frak a})^\ell\, (\frak a\,\widebar{\frak a})^p\,.
\end{multline*}
Since $\frak c$ is arbitrary in axiom \eqref{scale}, put
$\frak c=r\in\bbR$ for a moment. One obtains
\begin{equation*}
\smash{\sum_{\ell, p}}\gamma_{\ell p}\,\big\{r^{2p+\ell}-
\msf{const}\,(r)\big\}(\frak a+\widebar{\frak a})^\ell
(\frak a\,\widebar{\frak a})^p =0\qquad\forall\,
r,\frak a,\widebar{\frak a}
\end{equation*}
and, hence, nontrivial solutions for $\msf{const}\,(r)$ is possible
only if $2\,p+\ell$ is a fixed (external) integer; denote
it~$\textsf{K}$. Therefore, sum \eqref{N*} becomes the one of finitely
many terms and all of them are homogeneous in $\frak a$,
$\widebar{\frak a}$:
\begin{align}
N_{\!*}(\frak a,\widebar{\frak a})&=\sum\gamma_{\ell p}\,
(\frak a+\widebar{\frak a})^\ell\,
(\frak a\,\widebar{\frak a})^p\Big|_{2p+\ell=\textsf{K}}\notag\\
&=\sum_{p=0}^{\textsf{K}/2}\gamma_{\textsf{K}\sm2p,p}^{\mathstrut}\,
(\frak a+\widebar{\frak a})^{\textsf{K}\sm2p}\,
(\frak a\,\widebar{\frak a})^p\,.\label{K2m}
\end{align}

When $\textsf{K}=1,3,5,\ldots$ we have only the odd $(\textsf{K}-2\,p)$"=powers
$(\frak a+\widebar{\frak a})^1$, $(\frak a+\widebar{\frak a})^3$,
\ldots\ in the $p$-sum \eqref{K2m}. In such a case,
$N_{\!*}(\frak a,\widebar{\frak a})\sim (\frak a+\widebar{\frak a})$
and, hence, $N(\frak a)=0$ at $\frak a=\ri\,\bbR\ne0$. That
\textsf{K}"=case must be discarded because $N(\frak a)=0$ only if
$\frak a=0$ by the very statistical nature of the
$\frak a$"=coordinates. More formally, suppose the contrary, \ie, let
there exist some `specific' $\frak a'\ne0$ such that
$N(\frak a')=0$. From \eqref{scale} and \eqref{Na} there follows
\begin{gather*}
\forall\,\frak c{:}\quad
N(\frak c\,\frak a')=\msf{const}\,(\frak c)\times N(\frak a')=
\msf{const}\,(\frak c)\times 0=0\\
\hence\quad N(\frak c\,\frak a')=0\quad\hence\quad
N(\frak c')=0\quad\forall\,\frak c'\,;
\end{gather*}
the trivial solution.

Thus, only the even $\textsf{K}=0,2,4,\ldots$ and even powers
$(\textsf{K}-2\,p)\in\{\textsf{K}, \textsf{K}-2,\ldots,0\}$ of
$(\frak a+\widebar{\frak a})$ are allowed in \eqref{K2m}:
\begin{equation*}
N_{\!*}(\frak a,\widebar{\frak a})=
\gamma_0\,(\frak a+\widebar{\frak a})^0\,
(\frak a\,\widebar{\frak a})^\textsf{p}+
\gamma_2\,(\frak a+\widebar{\frak a})^2\,
(\frak a\,\widebar{\frak a})^{\textsf{p}\sm1}+\cdots
\end{equation*}
($2\,\textsf{p}\DEF\textsf{K}$). Homogeneity in $\frak a$ guides us, before
substituting this ansatz into \eqref{Ncc}, to switch over to the
modulus"=phase forms $\frak a=\varrho\,\re^{\ri \varkappa}$,
$\frak c=r\,\re^{\ri t}$:
\begin{equation*}
N_{\!*}(\frak a,\widebar{\frak a})=
\varrho^{2\textsf{p}}\,\{\gamma_0+\gamma_2\,{\cos^2}\varkappa+
\gamma_4\,{\cos^4}\varkappa+\cdots\}
\end{equation*}
(we renormalized $\gamma$'s). Then the scaling
$\frak a\mapsto\frak c\,\frak a$ amounts to the change
$(\varrho,\varkappa)\mapsto\,(\varrho\,r,\varkappa+t)$ in the latter
expression. One gets, instead of \eqref{Ncc},
\begin{multline*}
(\varrho\,r)^{2\textsf{p}}\,\big\{\gamma_0 +
\gamma_2\,{\cos^2}(\varkappa+t)+
\gamma_4\,{\cos^4}(\varkappa+t)+\cdots\big\}\\
=\msf{const}\,(r,t)\times\varrho^{2\textsf{p}}\, \{\gamma_0 +
\gamma_2\,{\cos^2}\varkappa+ \gamma_4\,{\cos^4}\varkappa+\cdots\}\,,
\end{multline*}
where all the variables $(\varrho,\varkappa;r,t)$ are understood to
be independent and equal in rights. It is immediately seen that
there is only one possibility here:
\begin{equation*}
\msf{const}\,(r,t)=\msf{const}'(r)=r^{2\textsf{p}}\,,\quad
\gamma_0=\text{free}\,,
\end{equation*}
and $\gamma_2=\gamma_4=\cdots=0$; put, for example, $\varkappa=0$.
As a result, only one term survives in sum \eqref{N*}:
\begin{equation*}
N(\frak a)=\gamma_0\times(\frak a\,\widebar{\frak a})^\textsf{p}
\end{equation*}
with yet free $\textsf{p}=1,2,3,\ldots$. Clearly, the
(\raisebox{-0.9ex}{\textasciitilde})"=involution \eqref{inv} would
yield the same answer:
\begin{multline}
\cal N\big[\frak a_1\bo\cdot\ket{\alpha_1} \h+
\frak a_2\bo\cdot\ket{\alpha_2}\h+\cdots\big]\\
=\msf{const}\times\big(|\frak a_1|^{2\textsf{p}}+|\frak a_2|^{2\textsf{p}}+
\cdots\big)\,.\label{ansatz}
\end{multline}

Getting ahead of ourselves, we could claim $\textsf{p}=1$ right here
because none of the values $\textsf{p}=2,3,\ldots$ may be preferable to
any other (the `world constant' $\textsf{p}\geqslant2$?), while $\textsf{p}=1$ is
minimal in this series. And yet, we address the device"=independence
\eqref{noncom} because it implies a changing of instruments
$\scr A\rightleftarrows\scr B$ and so the change of eigen-state
bases: $\big\{\ket{\alpha_1},\ket{\alpha_2},\ldots\big\}_{\!\!\!\scr A}
\rightleftarrows
\big\{\ket{\beta_1},\ket{\beta_2},\ldots\big\}_{\!\!\!\scr B}$.

When the family of $\scr A$"=distinguishable clicks coincides with
the family of the $\scr B$"=distinguishable ones
$\{\state\alpha_1,\ldots\} = \{\state\beta_1,\ldots\}$, we have actually
one and the same instrument: $\scr A=\scr B$. In the
$\ket\alpha$"=language, this means
\begin{equation*}
\big\{\ket{\alpha_1},\ket{\alpha_2},\ldots\big\}_{\!\!\!\scr A}=
\big\{\ket{\beta_1},\ket{\beta_2},\ldots\big\}_{\!\!\!\scr B}\,,
\end{equation*}
and the scale transformations
$\ket\alpha\mapsto\frak c\bo\cdot\ket\alpha=\ket\beta$ may be
disregarded here since the eigen"=states themselves\footnote{Notice
that the `eigen' does not mean here the eigen"=vector of an operator.
No operators appear in reasoning or in `$\bbH$"=mathematics' at the
moment. We also put for simplicity that the spectral labels assigned
to these vectors are non"=degenerated (\,$=$~dis\-tin\-guish\-able).} are
defined to within multiplicative constants. We then have to declare
transformations like
$\ket{\alpha_j}\mapsto\ket{\beta_s}\sim\ket{\alpha_k}$ as
\emph{trivial} permutations.

It is clear that the arbitrary permutation is formed from
transpositions like $\{\ket{\alpha_1}\mapsto\ket{\alpha_2}\,,
\ket{\alpha_2}\mapsto\ket{\alpha_1}\}$. Therefore it will suffice to
consider the 2"=dimensional changes and to exclude the trivial
diagonal (identical) and antidiagonal (transpositions) ones:
\begin{equation*}
\mbig[7](\!\!\!\begin{array}{l}
\ket{\beta_1}\\\ket{\beta_2}
\end{array}\![4]\mbig[7])=
\mbig[7](\!\!\begin{array}{cc}1&0\\0&1\end{array}\!\!\mbig[7])
\mbig[7](\!\!\!\begin{array}{l}
\ket{\alpha_1}\\\ket{\alpha_2}
\end{array}\![4]\mbig[7]),\;
\mbig[7](\!\!\!\begin{array}{l}
\ket{\beta_1}\\\ket{\beta_2}
\end{array}\![4]\mbig[7])=
\mbig[7](\!\!\begin{array}{cc}0&1\\1&0\end{array}\!\!\mbig[7])
\mbig[7](\!\!\!\begin{array}{l}
\ket{\alpha_1}\\\ket{\alpha_2}
\end{array}\![4]\mbig[7])\,.
\end{equation*}
The nontrivial basis"=changes, say, the simplest ones
\begin{equation}\label{kets}
\big(\ket{\alpha_1},\ket{\alpha_2};\ket{\alpha_3},\ldots\big)
_{\!\!\!\sss\scr A}\rightleftarrows
\big(\ket{\beta_1},\ket{\beta_2};\ket{\alpha_3},\ldots\big)
_{\!\!\!\sss\scr B}\,,
\end{equation}
correspond to observations by `non"=commuting devices'
$\scr A\ne\scr B$ and the latter do, without fail, exist in quantum
theory \cite[\textbf{\textsf{III}}-rd principium of \qm]{br}. We now have
to pass to the `erasing' the $\ket{{}\text{ket}}$"=symbols from
\eqref{star} because \eqref{Na} and the formal applying
\eqref{ansatz} to \eqref{star} ignore the down-arrow \eqref{noncom}
and thereby any relationships \eqref{kets} between $\ket\alpha$'s
and $\ket\beta$'s, as well as the very consequence \eqref{star}:
\begin{equation}\label{abab}
(\frak a_1\widebar{\frak a}_1)^\textsf{p}+
(\frak a_2\widebar{\frak a}_2)^\textsf{p}=
(\frak b_1\widebar{\frak b}_1)^\textsf{p}+
(\frak b_2\widebar{\frak b}_2)^\textsf{p}\,.
\end{equation}

Inasmuch as we have dealt with an \lvs-basis change \eqref{kets},
the coordinate representative $(\frak a_1,\frak a_2,\ldots)$ of (one
and the same) $\ket{{}\text{ket}}$"=vector \eqref{AB} undergoes an
associated \emph{linear} transformation $U$. In consequence, there
must exist the numeric changes
\begin{equation}\label{U}
\mbig[7](\!\!\!
\begin{array}{l}\frak a_1\\\frak a_2\end{array}\!\!\!\mbig[7])
\overset{U}{\mapsto}
\mbig[7](\!\!\!
\begin{array}{l}\frak b_1\\\frak b_2\end{array}
\!\!\!\mbig[7])=
\mbig[7](\begin{array}{cc}a&b\\c&d\end{array}\mbig[7])
\mbig[7](\!\!\!\begin{array}{l}\frak a_1\\\frak a_2\end{array}\!\!\!\mbig[7])
\end{equation}
and their (anti)diagonal subclass
\begin{equation*}
U=\mbig[7](\begin{array}{cc}a&0\\0&d\end{array}\mbig[7])\quad\text{or}\quad
U=\mbig[7](\begin{array}{cc}0&b\\c&0\end{array}\mbig[7])
\end{equation*}
should also be thought of as the trivial changes. Apart from the
obvious $\det U\ne0$, this yields the nontriviality condition for
\eqref{U}:
\begin{equation}\label{00}
a\,b\ne0\ne c\,d\,.
\end{equation}

That said, equality \eqref{abab} should be supplemented with
\eqref{U}--\eqref{00} and obeyed under all $\frak a$'s. Simplifying
notation $(\frak a_1,\frak a_2)\rightarrowtail(\bo x,\bo y)$, we
require
\def\xbar{\skew3\widebar{\itbf{x}}}
\def\ybar{\skew3\widebar{\itbf{y}}}
\begin{multline*}
(\bo x\,\xbar\,\,)^\textsf{p}+(\bo y\,\ybar\,\,)^\textsf{p}\\
=(a\,\bo x+b\,\bo y)^\textsf{p}
(\widebar{a\,\itbf{x}+\mit b\,\itbf{y}}\,\,)^\textsf{p}+
(c\,\bo x+d\,\bo y)^\textsf{p}
(\widebar{c\,\itbf{x}+\mit d\,\itbf{y}}\,\,)^\textsf{p}
\end{multline*}
for all $(\bo x, \xbar\,, \bo y, \ybar\,\,)$, which are understood to
be independent variables. By expanding, some binomial expansions
arise ($\textsf{p}\geqslant2$):
\begin{multline*}
\bo x^\textsf{p}\,\xbar^\textsf{\,\,\,p}+\bo y^\textsf{p}\,\ybar\,^\textsf{\,\,p}=
(a^\textsf{p}\,\widebar a^\textsf{p}+c^\textsf{p}\,\widebar c^\textsf{\,\,p})\cdot
\bo x^\textsf{p}\,\xbar^\textsf{\,\,\,p}+\cdots\\
+\textsf{p}^2\cdot\big\{(a\,\bo x)^{\textsf{p}\sm1}\,(b\,\bo y)\cdot
(\widebar{a\,\itbf{x\,}}\,)^{\textsf{p}\sm1}\,
(\widebar{b\,\itbf{y}}\,\,)\qquad\qquad\,\\\quad\,\,\,+
(c\,\bo x)^{\textsf{p}\sm1}\,(d\,\bo y)\cdot
(\widebar{c\,\itbf{x\,}}\,\,)^{\textsf{p}\sm1}\,
(\skew1\widebar{d\,\itbf{y}}\,\,)\big\}+\cdots\\
+(b^\textsf{p}\,\overline{b}\,^\textsf{p}+d\,^\textsf{p}\,
\skew1\widebar{d}\,^\textsf{p})\cdot\bo y^\textsf{p}\,\ybar^\textsf{\,\,\,p}=
\cdots\,,
\end{multline*}
where only one cross-term $(\bo x\,\xbar\,\,)^{\textsf{n}}
{\cdot}\,(\bo y\,\ybar\,\,)^{\textsf{m}}$ has been displayed. Collecting in
$\bo x\,\xbar$ and $\bo y\,\ybar\,$, one gets (among other terms)
\begin{multline*}
\cdots=\cdots+\textsf{p}^2\cdot\big\{
\uwave[5]{\text{$|a|^{2\textsf{p}\sm2}\,|b|^2+
|c|^{2\textsf{p}\sm2}\,|d|^2$}}\big\}\\ \times
(\bo x\,\xbar\,\,)^{\textsf{p}\sm1}\, (\bo y\,\ybar\,\,)+ \cdots\,.
\end{multline*}
Clearly, such expressions have always been present in the sum and
the wavy"=emphasized term must be zero. Hence,
\begin{equation*}
|a^{\textsf{p}\sm1}\,b|^2+|c^{\textsf{p}\sm1}\,d|^2=0 \quad\hence\quad
\{a\,b=0=c\,d\}\,.
\end{equation*}
This contradicts \eqref{00}. Only trivial permutations
(consequently, only `commuting devices') are allowed under
$\textsf{p}\geqslant2$. Thus, $\textsf{p}=1$ and \eqref{ansatz} is refined:
\begin{multline}\label{sum2}
\SL\,\big(\frak a_1\bo\cdot\ket{\alpha_1} \h+
\frak a_2\bo\cdot\ket{\alpha_2}\h+\cdots\big)\\
=\msf{const}\times\big(|\frak a_1|^2+|\frak a_2|^2+ \cdots\big)\,,
\end{multline}
where $\msf{const}$ must be a common, while free, constant for all
the $\ket\alpha$"=representations. Extended to the higher dimensions
$(N\times N)$, the $U$"=matrices \eqref{abab}--\eqref{U} preserve the
sum of squares \eqref{sum2}, and matrix $(U^{\sss\top})^{\sm1}$
determines a corresponding change of the base vectors
$\ket{\alpha_j}$ into the other ones. Thereby, an abstract
$\oper[-2]{\cal U}$"=transform on $\bbH$ has been well defined, and
it is known to be nontrivial. Call this property \emph{unitarity},
and that's the point where this concept comes into quantum theory.

We now return to the task \eqref{nu}. The semantics \eqref{infty}
suggests the only way of harmonizing the `theoretical
infinity~$\infty$'---$\Sigma$"=postulate---with finite
quantities\footnote{Hilbert: ``Das Operieren mit dem Unendlichen kann
nur durch das Endliche gesichert werden''.} coming from experiment;
their $\bbR^{\sss+}$"=numerical images, to be precise. Namely, we
introduce by definition the concept (it was not so far) of the
micro-events' long-run frequencies:
\begin{equation*}
\nu_k\DEF\frac{\SL\,_k\times\infty}{\sum_j (\SL\,_j\times\infty)}\,.
\end{equation*}
Finally, the completed formulation of Born's result has not been
exhausted by the squares' formula.
\begin{itemize}
\item \embf{The $\bo{2}$-nd theorem of quantum
 empiricis\hbox to0ex{m}.}
\begin{enumerate}
\item[1)] \emph{Basis"=independence}: the sum of squares \eqref{sum2}
 is the only rule that is compatible with the \SL"=additivity and the
 `device non"=commutativity'
 $\big\{\ket{\alpha_j}\big\}_{\!\!\sss\scr A}\ne
 \big\{\ket{\beta_k}\big\}_{\!\!\sss\scr B}$.

\item[2)] \emph{The $\oper[-2]{\cal U}$"=equivalence of bases}: the
 changing of observational instruments
 $\scr A\rightleftarrows\scr B$ is represented in $\bbH$ by unitary
 transformation $\big\{\ket{\alpha_j}\big\}
 \overset{\raisebox{-0.2ex}{$\scs\smash[b]{U}$}}{\rightleftarrows}
 \big\{\ket{\beta_k}\big\}$ between their eigen"=states.

\item[3)] The $\state\alpha$-events' statistics for representation
 \eqref{=} is approximated according to \emph{the Born rule}
 \begin{equation}\label{rule}
 \nu_k=\frac{|\frak a_k|^2}{|\frak a_1|^2+|\frak a_2|^2+\cdots}\,.
 \end{equation}
\item[4)] No use is required of the
 Hilbertian/""ten\-sor/""orthogonality/""projec\-tor%
 /""ope\-ra\-tor/""\ldots/""uni\-tarity structures when deducing the rule.
\end{enumerate}
\end{itemize}

The rudimentary physics at the moment is just the click collections.
Therefore the rule \eqref{rule} does not require---it should also be
emphasized---any physical terminology: interactions, dynamics,
evolution, measuring processes, apparatus, \etc. Nor does the
derivation address a density matrix---mixture of
$\ket\alpha$'s---and such concepts as space/""time/""causality (in the
\textsc{epr}"=controversy, say), (non)relativity, gravity\footnote{In
particular, the binding the rule to unitarity or $t$"=dynamics would
be contradictive and entail a grave problem of reconciliation with
the well-known issues in quantum gravity \cite{kuchar}: the problem
with the very Hilbert (and Fock) space, with the dynamical
(non"=fixed as in \qft) background and the observer"=dependent
concepts of particles and their number, of time itself, \etc.}, and
(non)inertial reference frames; to say nothing of the moot and
debatable \cite{saunders, rae, wallace} notions like collapses, `the
world(s)/mind(s)', the MWI"=bifurcations of the universe
\cite{saunders, everett}, (classical/""objective) reality, or a
subjective/""anthropic \cite[p.~155--165]{dewitt} category of the
rational belief/""preference \cite{saunders}. In essence, we have made
do only with the two obvious premises: \eqref{N+} and \eqref{star}.
These are obligatory requirements, which is why the word \SL\ may be
formally even cast away from the theorem. The quadratic dependence
above is, roughly speaking, a mathematical statement concerning the
correctly defined---invariance \eqref{star}---function on $\bbH$
with $\ket\alpha$"=additivity \eqref{N+}. An additive property, in
one form or another, is present almost in all works on the rule
\cite{deutsch, dewitt, gleason, gudder, assis, hartle, saunders1,
busch, saunders, zela}, and each derivation of formula \eqref{rule}
is in anyway a construction of a map from more primary tenets, even
though we do not pronounce explicitly (or subconsciously) the
thesis~(\hyperlink{I}{$\bullet$}).

It is also clear that the proposed inference procedure appears to be
unique---no non-Born statistics exists---because the premises
\eqref{N+} and \eqref{noncom}--\eqref{star} are `non"=reducibly'
minimal; as it should be for a formula, which by itself defines the
math of a quantum-lab experiment at the very low level, \ie, the
meaning of the words `handling the click collections'.

\section{Discussion}

\subsection{How would derivation look in the orthodoxy?}

Let us forget the \SL"=conception and theorem about linearity of
quantum superposition---`accumulation of clicks into coefficients
$\frak a_j$' \cite{br}, and the very doctrine of numbers in \qm.
What points should be introduced into the quantum axiomatics in
order to derive \eqref{rule}?

First of all, we should accept the statistical treatment of the
$\frak a_j$"=coordinates. It is widely known as early as the 1926
works by Born himself \cite{born}. The words
``Statistik/""statistischen'' appear at the very end of the first brief
communication \cite{born} (and disappear in the second of the works
\cite{born}); though in the context of the particle"=collision
processes, \emph{not of the abstract} micro"=events.

The (relative) frequency view of the state-rays in a Hilbert
space---the multiplicatively statistical reading of the equivalence
$\ketPsi\approx\frak c\bo\cdot\ketPsi$---suggests to give up the
notions like `up to a constant, inaccessible phases, \etc' and to
deal with the non"=normalized ($\h+$)"=sums\footnote{Whether this idea
has been expressed in the literature, the author is not aware. I
would be grateful for an information in
this regard.} %
\makeatletter\def\@makefnmark{\textup{\red*}}\makeatother%
\eqref{=}, \ie, without constraint
$\vcenter{\Larger{\|}}\ketPsi\vcenter{\Larger{\|}}=1$;
cf.~\cite[p.~185]{dewitt}. The ($\bo\cdot$)"=normalization and
($\h+$)"=summation are the opposing requirements but linearity is of
course primary. (Parenthetically, the math"=normalization of a
\qm"=state has nothing to do with its statistical nature). Therefore,
a certain notion of the \emph{additive} `quantifying/""sizing' must be
introduced \cite[p.~1296]{zela}. Such an additivity manifests in the
well-known orthogonality and distinguishability of eigen"=states.
See, \eg, \cite{zurek2, gleason}, page~890 in \cite{assis}, and also
a concept of the orthogonal additivity in \cite[sect.~5.2]{gudder}.
The $\state\alpha$-, $\ket\alpha$"=distinguishability is thus of
fundamental importance when deducing both the \lvs\ and \SL\
structures.

Further, the difference between an abstract state (or an abstract
sum $\frak a\bo\cdot\ketPsi\h+\frak b\bo\cdot\ketPhi$) and its
$\bbC$"=numerical basis"=dependent $\ket\alpha$"=representation (in a
reference frame for the `observer $\scr A$') does of course not go
away and remains the conceptual point \cite[sect.~8.3]{br}. No
$\scr A$"=instrument is exclusive because any
$\ket\alpha$"=preference---\eg, privileged observables or pointer
states \cite{saunders} in some takes on the `measurement
problem'---would run counter to the basic principle of the
representation invariance of physical theories and of
\qm"=mathematics \eqref{noncom} in particular; ``democracy of bases'',
by Jeffrey Barrett.

We should also declare what the complex $(*)$"=conjugation does in
\qm"=theory; except for a scalar"=product axiom. The declaration is
this: a $(*)$"=invariance of the $\nu$"=statistics. Subsequent
actions, including the reading of the device/""operator
non"=commutativity $\scr A\ne\scr B$, do not then require any
postulations and have been described in the previous section. The
sequence $\eqref{times} \rightarrowtail \eqref{axiom}
\rightarrowtail \eqref{Cx}$ and point~4) in the theorem remain in
force.

Informally, to disclose `Born's square' by manipulating the
$\ket{{}\text{ket}}$"=symbols like
$\oper[-3]{\msf P}_{\!\!\!\ket\alpha}\ket{{}\psi}$ is a rather
non"=efficient way, to say the least. The Born rule is a statement
not about $\ket{{}\texttt{in}}{\dashrightarrow}\ket{{}\texttt{out}}$
reductions or von~Neumann's $\oper[-3]{\msf P}$"=projectors but about
numbers. That is, about $\bbC$"=representatives
$(\frak a_1,\frak a_2,\ldots)$ irrespective of their calculation method
$\frak a=\langle\alpha|\psi\rangle$, because
\emph{$\frak a$"=coefficients are not `aware of' the binary
structures on \lvs;} neither of the inner"=product nor of the
orthogonality \cite{reed}. Expressed differently, rather than being
a consequence of the ideology $\lceil$Hilbert space $+$ physical
con\-sid\-er\-a\-tion$\rceil$, the rule does determine the Hilbert
structure itself; a definitional superstructure over \lvs.

A.~Gleason (with his famous representation theorem \cite{gleason})
and H.~Everett \cite{everett} were perhaps the first to attempt at
vindicating the rule in the framework of the orthodox axiomatics.
Everettian approach came under criticism of many authors
\cite{ballentine, mandolesi, saunders, rae} and later N.~Graham
\cite{graham} and J.~Hartle \cite{hartle} reconsidered Everett's
conclusions through the frequency operator as an observable; see,
however, \cite{squires}.

\begin{comment}
When deducing the rule, Everett \cite[p.~71]{everett} freely changes
the function arguments, puts ``$\frak M(\texttt{a}_\texttt{i})=
\frak M(\sqrt{\texttt{a}_\texttt{i}^\texttt{*}\,\texttt{a}_\texttt{i}^{}})$'' and
does ``impose the additivity requirement'', then restricts ``the choice
of $\frak M$ to the square amplitude'' and puts
``$\frak M(\texttt{a}_\texttt{i}^{})=
\texttt{a}_\texttt{i}^\texttt{*}\,\texttt{a}_\texttt{i}^{}$'', does ``replace the
$\texttt{a}_\texttt{i}^{}$ by their amplitudes
$\mu_\texttt{i}=|\texttt{a}_\texttt{i}^{}|$'', defines ``a new function
$\text{g(x)}=\frak M(\sqrt{\text{x}})$'', etc, etc. Finally, on
p.~72, he draws a conclusion that ``the only choice \ldots\ is the
square amplitude measure''. That is to say, by use of the fact that
square of a coefficient is a sum of other squares (Hilbert), one
infers a rule of squares. Clearly, in no way is this any proof
\cite[p.~236]{graham}, \cite[pp.~163, 185]{dewitt}, however, its
`refinements and justifications' have got even into textbooks
\cite[sect.~8.4.1, ``Everett's theorem'']{ivanov}.
\end{comment}

It is also not clear, what would be changed in reasoning on
pp.~71--72 of \cite{everett}, if the two and
\raisebox{.7ex}{$\sqrt[\uproot{2}2]{}$} would be substituted for
$\textsf{p}$ and \raisebox{.7ex}{$\sqrt[\uproot{3}\textsf{p}]{}$}.
Expressed another way, why and which the $L^{\!\!\textsf{p}}$-norms are
relevant to the quantum state"=space?

\subsection{Math-rigors: topology, continuity, and the like}

The latter question was fully considered by S.~Aaronson in the work
\cite{aaronson} wherein the exclusiveness of an $L^{\!\!2}$-norm was
justified. His analysis, besides other important questions, is
extended even to non"=integer \textsf{p}'s, and realization of
device"=independence $\scr A\rightleftarrows\scr B$ by the
$U$"=matrices above fits completely Aaronson's idea of the (power
dependence) norm's preservation under linear transformation. In this
context, the Pythagorean theorem Aaronson mentions \cite[pp.~2,
4]{aaronson} should be thought of as the only possible way of
introducing the very first numeric quantity in quantum theory---the
function~\eqref{sum2}. In other words, the \qm"=version of
Pythagoras' theorem is not a theorem, and even not a
$\bbC$"=orthogonality property $\ket\alpha\bot\ket\beta$ \cite{reed},
but \emph{merely a definition} \cite{brH} of additivity
\begin{equation*}
\cal N\big[\frak a\bo\cdot\ket\alpha\h+
\frak b\bo\cdot\ket\beta\big]=\cal N[\frak a\bo\cdot\ket\alpha]
+\cal N[\frak b\bo\cdot\ket\beta]\,,
\;\forall\frak a,\frak b\in\bbC
\end{equation*}
in the language of vector $\bbH$"=representatives $\ket\alpha,
\ket\beta$ to the distinguishable (\,$=$ eigen) quantum micro"=events
$\state\alpha \not\approx \state\beta$. In the accustomed notation
for `lengths', and with Born's square, this additivity might be
written as
\begin{equation*}
\vcenter{\Larger{\|}}\frak a\bo\cdot\ket\alpha\h+
\frak b\bo\cdot\ket\beta\vcenter{\Larger{\|}}^2=
\vcenter{\Larger{\|}}\frak a\bo\cdot\ket\alpha\vcenter{\Larger{\|}}^2
+
\vcenter{\Larger{\|}}\frak b\bo\cdot\ket\beta\vcenter{\Larger{\|}}^2
\,,\quad\forall\frak a,\frak b\in\bbC\,.
\end{equation*}

On the other hand, the state"=space $\bbH$ is almost a `bare' \lvs\
at the moment. It is neither a normed nor a topological space
\cite{reed}, because construction of (continuous) maps from
$\bbH$---no matter where---does not yet arise as a task. Inasmuch as
the states themselves are not observable entities (whatever that
means \cite[sect.~10]{br}) and are not yet comparable with each
other, the low level quantum `$\bbH$"=mathematics' does not care
questions like `whether we need a construction
\hbox{$\|{\cdot{\cdot}\cdot}\|$} with axioms of a norm?---the
triangle inequality, \etc'. The more so as there is an equivalence
relation on norms in the finite"=dimensional \lvs\ \cite{reed}; \eg,
the $L^2$"=norm is (topologically) equivalent to the $L^1$"=norm.
Since \qm"=theory needs to be a quantitative one, the topology should
be implemented in a numeric way, \ie, through $\frak a$'s.

The \qm"=empiricism in turn does not yet give grounds to introduce
any functions on $\bbH$, other than \SL. We thus draw a conclusion
that if such a function is exclusive, that is how it will induce the
topology on the abstract $\bbH$"=vectors by means of the numeric
$\cal N$"=function \eqref{sum2} of their
$\ket\alpha$"=representatives. This point does precede the Hilbert
space and Born statistics, and not the other way round. The quantum
state"=space can thus be turned into an $L^2$"=normed vector space
whose topology conforms to the $\bbC$- and $\bbR$"=field topology of
numbers~$\frak a_j$.

This $\bbR$"=topology has already been used when deriving the
\textsf{C}"=function \eqref{Cx}\footnote{All the other solutions to this
equation are ``pretty `weird'\,\,'' (J.~Acz\'el--J.~Dhombres). They are
globally/""locally irregular \cite[pp.~129--130]{kuzma} and their
graphs are everywhere dense in $\bbR^2$.}. On the other hand, ansatz
\eqref{N*} should be understood not as the (infinite) series in
$(\frak a, \frak a^*)$ but just as a finite (purely algebraic)
symmetrical sum. Otherwise, if this were the `infinity'"=case, we
would deal with a non"=motivated non"=algebraic extension of the
`pure' $\bbH$"=algebra and thereby with some extra"=topological
requirements that do not follow from empiricism. However, the
restriction on such an `implied infinity' is not a loss of
generality because, in any case, homogeneity
\eqref{Ncc}--\eqref{K2m} extracts the only term from~\eqref{N*}.

Yet a further aspect of function $\cal N$ concerns the very
statement of the problem. Every \lvs\ has infinitely many bases.
However, as the space $\bbH$ was arising alongside the bases of
observables \cite{br}---eigen"=vectors $\ket{\alpha_j}$, let us ask
ourselves the question: What is the way in which the basis of an
observable stands out from the other abstract bases, which are as
good as any one? Quantum empiricism tells us that all one has to do
is to invoke some statistical considerations. These will boil down
to the following semantic supplement: a certain function on $\bbH$
(better to say, functional), which reflects the natural"=language
notion of the accumulating the distinguishable micro"=events; \ie,
$\lceil$ad\-di\-ti\-vi\-ty $=$ mutual ex\-clu\-si\-v\-ity$\rceil$. The
presence (or non) of such a numeric function---a new math (scalar)
add-on over $\bbH$---will determine these `good bases'. Thus the
mathematics accompanying the quantum statistics---motivation and the
\textbf{Definition} itself---can be restated as a question of special
bases of \lvs\ and has a quite minimalistic formalization:

\begin{itemize}
\item[$\diamondsuit$]Given an \lvs\ of quantum states, define the
 $\scr A$"=base(s)---due to \qm"=non"=commutativity, it must be not
 unique---by the following requirement. Basis
 $\big\{\ket{\alpha_j}\big\}$ is referred to as \emph{basis of an
 observable} $\scr A$ if there exists a well"=defined function(al)
 $\cal N$ on $\bbH$, which satisfies the properties of
 $\ket\alpha$"=additivity \eqref{N+} and of $(*)$"=invariance
 \eqref{Ninv} for all $\frak a$'s.
\end{itemize}

Is such a definition meaningful? What is the function? How is it
derived? Whether it exists (in what sense?) and is
unique/""nontrivial? What is relationship between different
$\scr A$"=bases? A group, the $\cal U$"=equivalence class? Where does
it come from? The answers to these questions are the derivation of
the theorem from \eqref{N+}--\eqref{Ninv}. All the other
bases---beyond the unitarity condition---remain the abstract ones in
\lvs. In particular, that such a function is known to exist for
certain bases follows, again, from the fact that the $\bbH$"=space
itself was being fabricated from the statistical
$\ket\alpha$"=representatives. This point---pt.~4) in the 1-st
Theorem \cite{br}---elucidates what should be meant by the ``almost a
`bare' \lvs'' above. The $\bbH$"=space is not a completely abstract
\lvs, but one that must have been equipped with a numeric
superstructure $\cal N$. Parenthetically, the same method provides a
tool of deriving the `topological $\cal N$"=function' for other
linear manifolds: different numeric fields, different involutions,
\etc.

Summing up, the questions of topology on the $\bbH$"=space (and on
numbers) are, strictly speaking, to be solved simultaneously with
the construction of function $\cal N$, which, in turn, comes from
quantum empiricism as the \SL. There is also no difficulty in
extending the above theorem to mixtures of states---the Gleason
theorem \cite{gleason}, when using the orthodox \qm"=mathematics of
operators.

Of course, the reasoning given in this subsection is not quite
rigorous arguments and is merely a mathematical ideology. However we
are of the opinion that the entire quantum foundations, and not just
their algebraic \lvs"=constituent, admit a considerable strengthening
the mathematical motivation and rigor---a proposal for the
mathematics experts---even to the extent of pedantic justification
of all the topologies/""ordering, of the (general quantum) case $\dim
\bbH=\infty$, of a numeric domain--- $\bbC$, $\bbR$, or
$\bbR^{\sss+}$---and the like. (Of course, wouldn't have to presume
the positive values for \SL, as the $\bbR^{\sss+}$"=domain came on
its own.) In the first place, this fully applies to the work
\cite{br}. The more so as the mathematical grounds to the semantic
notions of continuity, connectivity, and the physical (numeric)
lexicon of completeness, approximations, limits, infinitesimal
$\eps$'s, convergence, \etc\ have long been formalized in topology
\cite{reed, kuzma}.

\subsection{Remarks on spacetime}

A word on the physical $(3\,{+}\,1)$"=spacetime. This topic bears on
the full $(\bo x,t)$"=representation---the continuous
$(\bo x,t)$"=parameters of automorphism---of the invariant
$\bbH$"=theory of $\lceil\scr A$"=bases $+$ $\ket{{}\text{ket}}$'s $+$
Born's uni\-tar\-ity$\rceil$, because the abstract states themselves
are not tied to the chrono"=geometrical notions of
causality/""(non)lo\-cal\-ity/""propagation"=speed (of something, say of
light $c$) and to ``the objective determination of space"=time
phenomena'' (W.~Heisenberg). Here, there is no way to bypass the
matters of principle. Among them: a state"=space separability,
accurate introduction of the $\bbH$"=representatives to observables
in \qm/\qft\ and of Hilbert's space itself; why/""where the binary~(?)
inner~(?) product~(?) on $\bbH$ comes from \cite{brH}. (Realization
of the Hilbert space in quantum gravity is not a `$t$"=constant', as
with the elementary \qm.) We should also ascertain ``what is to be
regarded as an observable in a quantum theory of gravity''
\cite[pp.~107, 91]{isham}, \cite{kuchar} and the numeric labeling
the space"=time continuum, degrees of freedom, dimension $D=1+3$ (?),
and other data. In particular, we encounter non"=rhetorical question
about bringing formula \eqref{rule} into correlation with
non"=discrete (and conventional) constructions like
\begin{equation*}
\smash{\big|\psi(x)\big|^2\,dx\,,\qquad\big|\psi(x,t)\big|^2\,dx\;\;(?)\,.}
\end{equation*}

The global inference to be drawn from these remarks and (marked)
questions is such that the coherent strategy needs to be not a
relativistic \qft"=generalization of \qm\ followed by a quantization
of the gravity (to be renormalizable?)\ `as of fields', but a direct
creation of a framework for entirely covariant theory, in which all
the $\lceil$math $+$ phys$\rceil$"=ingredients---clicks' statistics
and quantum nature of observables/""spectra/""coordinates on manifolds
(the equivalence principle)---are consistently introduced at the
level of all possible $(\bo x, t)$"=realizations to the abstract
$\bbH$"=space ab~initio.

These (difficult) matters call for special consideration, and (some
of them) will be treated at length elsewhere. The absence of the
word `probability' in the present work is no accident
\cite[sect.~11.2]{br}. As we have seen, the micro"=events
supplemented with the \lvs"=structure---superposition principle---do
not require such a concept.

The author would like to thank the QFT"=department staff of TSU for
discussions. The work was supported by a grant of Tomsk State
University.


\begin{thebibliography}{99}
\def\showkeysBib{l}

\bibitem{aaronson} \textsc{Aaronson~S.}
\emph{Is Quantum Mechanics An Island In Theory\-space?}
\href{https://arxiv.org/abs/quant-ph/0401062}%
{\blue\texttt{https:/\!\!\!/arxiv.org/abs/quant-ph/}}\\
{\blue\texttt{0401062}} (2004), 1--9.


\bibitem{assis} \textsc{Assis~A.~V.~D.~B.}
\emph{On the nature of $\pow{a}{k}{*}a_k$ and the emergence of the
Born rule.} Ann. Phys. (Berlin) (2011) \textbf{523}(11), 883--897.

\bibitem{ballentine} \textsc{Ballentine~L.~E.}
\emph{Can the Statistical Postulate of Quantum Theory be
Derived?---A Critique of the Many"=Universes Interpretation.} Found.
Phys. (1973) \textbf{3}(2), 229--240.


\bibitem{born} \textsc{Born~M.}
\emph{Zur Quantenmechanik der Sto\ss{}vorg\"ange.} Zeit. Phys.
(1926) \textbf{XXXVII}, 863--867; \textbf{XXXVIII}, 803--827.


\bibitem{br} \textsc{Brezhnev~Yu.~V.}
\emph{Linear superposition as a core theo\-rem of quantum
empiricism.} \href{https://arxiv.org/abs/1807.06894}%
{\blue\texttt{https:/\!\!\!/arxiv.org/abs/}}\\
{\blue\texttt{1807.06894}} (2018), 1--70.

\bibitem{brH} \textsc{Brezhnev~Yu.~V.}
\emph{Whence the Hilbert space?} In preparation.

\bibitem{busch} \textsc{Busch~P.}
\emph{Quantum States and Generalized Observables: A Simple Proof of
Gleason's Theorem.} Phys. Rev. Lett. (2003) \textbf{91}(12), 120403(4).

\bibitem{caves} \textsc{Caves~C.~M.}
\emph{Notes on Zurek's derivation of the quantum
probability rule.} \href{%
http://info.phys.unm.edu/~caves/reports/ZurekBornderivation.pdf}%
{\blue\texttt{http:/\!\!\!/info.phys.unm.edu/%
\raisebox{-0.7ex}{\textasciitilde}caves/}}\\
{\blue\texttt{reports/ZurekBornderivation.pdf}} (2005), 1--6.

\bibitem{deutsch} \textsc{Deutsch~D.}
\emph{Quantum theory of probability and decisions.} Proc. Royal Soc.
London~\textbf{A} (1999) \textbf{455}, 3129--3137.

\bibitem{dewitt} \textsc{DeWitt~B.~S. \& Graham~N. (\textup{eds})}
\emph{The Many-Worlds Interpretation of Quantum Mechanics.}
Princeton University Press (1973).

\bibitem{everett} \textsc{Everett~H., III.}
\emph{The Theory of the Universal Wave Function.} PhD~Thesis,
Princeton University (1956). In \cite{dewitt}, pp.~3--140.

\bibitem{gleason} \textsc{Gleason~A.~M.}
\emph{Measures on the Closed Subspaces of a Hilbert Space.} Journ.
Math. Mech. (1957) \textbf{6}(6), 885--893.

\bibitem{graham} \textsc{Graham~N.}
\emph{The measurement of relative frequency.} In \cite{dewitt},
pp.~230--253.

\bibitem{gudder} \textsc{Gudder~S.~P.}
\emph{Stochastic Methods in Quantum Mechanics.} North Holland
(1979).

\bibitem{hartle} \textsc{Hartle~J.~B.}
\emph{Quantum mechanics of individual systems.} Amer. Journ. Phys.
(1968) \textbf{36}(8), 704--712.

\bibitem{isham} \textsc{Isham~C.~J.}
\emph{Canonical Quantum Gravity and the Problem of Time.}
\href{https://arxiv.org/abs/gr-qc/9210011}%
{\blue\texttt{https://arxiv.org/abs/gr-qc/9210011}} (1992), 1--124.

\bibitem{jarlskog} \textsc{Jarlskog~C.}
\emph{A Comment on a Recent Derivation of the Born Rule by Zurek.}
\href{https://arxiv.org/abs/1107.3736}%
{\blue\texttt{https:/\!\!\!/arxiv.org/abs/1107.3736}} (2011), 1--4.


\bibitem{kuchar} \textsc{Kucha\v{r}~K.~V.}
\emph{Time and interpretations of quantum gravity.} Int. Journ. Mod.
Phys. \textbf{D} (2011) \textbf{20}(1), 3--86.

\bibitem{kuzma} \textsc{Kuczma~M.}
\emph{An Introduction to the Theory of Functional Equations and
Inequalities.} Birkh\"auser (2009).

\bibitem{mandolesi} \textsc{Mandolesi~A.~L.~G.}
\emph{Analysis of Wallace's proof of the Born rule in Everettian
quantum mechanics II: Concepts and axioms.} Found. Phys. (2019)
\textbf{49}(1), 24--52.

\bibitem{masanes} \textsc{Masanes~L., Galley~T.~D. \& M\"uller~M.~P.}
\emph{The measurement postulates of quantum mechanics are
operationally redundant.} Nature communications (2019)
\textbf{10}(1361), 1--6.

\bibitem{rae} \textsc{Rae~A.~I.~M.}
\emph{Everett and the Born Rule.} Stud. Hist. Phil. Mod. Phys.
(2009) \textbf{40}(3), 243--250.

\bibitem{reed} \textsc{Reed~M. \& Simon~B.}
\emph{Methods of modern mathematical physics $\bo1$. Functional
analysis.} Academic Press, Inc (1980).

\bibitem{saunders1} \textsc{Saunders~S.}
\emph{Derivation of the Born Rule from Operational Assumptions.}
Proc. Royal Soc. London~\textbf{A} (2004) \textbf{460}, 1--18.

\bibitem{saunders} \textsc{Saunders~S., Barrett~J., Kent~A. \& Wallace~D.}
\emph{Many worlds? Everett, Quantum Theory, and Reality.} Oxford
University Press (2010).

\bibitem{sinha+} \textsc{\hbox{Sinha~U., Couteau~C., Jennewein~T.,
Laflamme~R.} \& Weihs~G.} \emph{Ruling Out Multi-Order Interference
in Quantum Mechanics.} Science (2010) \textbf{329}, 418--421.

\bibitem{sinha} \textsc{Sinha~U., Couteau~C., Medendorp~Z., Sollner~I.,
Laflamme~R., Sorkin~R. \& Weihs~G.} \emph{Testing Born's Rule in
Quantum Mechanics with a Triple Slit Experiment.} AIP Conf. Proc.
(2009) \textbf{1101}, 200--207.

\bibitem{squires} \textsc{Squires~E.~J.}
\emph{On alleged ``proof'' of the quantum probability law.} Phys.
Lett. \textbf{A} (1990) \textbf{145}(2/3), 67--68.

\bibitem{wallace} \textsc{Wallace~D.}
\emph{Everettian Rationality: Defending Deutsch's Approach to
Probability in the Everett Interpretation.} Stud. Hist. Phil. Mod.
Phys. (2003) \textbf{34}(3), 415--439.

\bibitem{zela} \textsc{De~Zela F.}
\emph{Gleason-Type Theorem for Projective Measurements, Including
Qubits: The Born Rule Beyond Quantum Physics.} Found. Phys. (2016)
\textbf{46}(10), 1293--1306.

\bibitem{zurek1} \textsc{Zurek~W.~H.}
\emph{Environment"=Assisted Invariance, Entanglement, and
Probabilities in Quantum Physics.} Phys. Rev. Lett. (2003)
\textbf{90}(12), 120404(4).

\bibitem{zurek2} \textsc{Zurek~W.~H.}
\emph{Probabilities from entanglement, Born's rule
$p_k^{}=|\psi_k^{}|^2$ from envariance.} Phys. Rev.~\textbf{A} (2005)
\textbf{71}(5), 052105(29).


\bibitem{zurek3} \textsc{Zurek~W.~H.}
\emph{Quantum Theory of the Classical: Quantum Jumps, Born's Rule,
and Objective Classical Reality via Quantum Darwinism.} Phil. Trans.
Royal Soc.~\textbf{A} (2018) \textbf{376}, 20180107(26).

\bibitem{ivanov} \textsc{Èâàíîâ~Ì.~Ã.} \emph{Êàê ïîíèìàòü êâàíòîâóþ ìåõàíèêó?}
Ðåãóëÿðíàÿ è õàîòè÷åñêàÿ äèíàìèêà (2015).\\
\href{https://mipt.ru/students/organization/mezhpr/%
biblio/q-ivanov.php}%
{\blue\texttt{https:/\!\!\!/mipt.ru/students/organization/mezhpr/}}\\
{\blue\texttt{biblio/q-ivanov.php}}

\end{thebibliography}
\end{document}